\documentclass[journal]{IEEEtran}

\usepackage{graphicx,multicol}
\usepackage{psfrag,amsbsy,pstricks,amsmath,amsfonts,amssymb,cite}
\usepackage{algorithm,algpseudocode,enumerate}
\usepackage{epstopdf}

\usepackage{color}
\usepackage{graphicx}

\usepackage{booktabs}
\DeclareMathOperator*{\argmin}{arg\,min}
\newcommand{\GA}{{\mathrm G}}

\newcommand{\x}{\boldsymbol{x}}
\newcommand{\s}{\boldsymbol{s}}
\newcommand{\e}{\boldsymbol{e}}
\newcommand{\0}{\boldsymbol{0}}
\newcommand{\I}{\boldsymbol{I}}
\newcommand{\V}{\boldsymbol{V}}
\newcommand{\h}{\boldsymbol{h}}

\newcommand{\m}{\boldsymbol{m}}
\newcommand{\n}{\boldsymbol{n}}
\newcommand{\rr}{\boldsymbol{r}}
\newcommand{\bb}{\boldsymbol{b}}
\newcommand{\cc}{\boldsymbol{c}}

\newcommand{\G}{\boldsymbol{G}}

\newcommand{\tra}{^{\textrm{T}}}
\newcommand{\md}{\textrm{M}}

\newcommand{\gpdf}{\mathrm{N}}
\newcommand{\dd}{\,\mathrm{d}}
\newcommand{\subs}[1]{[{#1}]}

\newcommand{\MA}{{\cal X}}

\newcommand{\rB}{\mathrm{B}}
\newcommand{\rC}{\mathrm{C}}
\newcommand{\rO}{\mathrm{O}}
\newcommand{\rT}{\mathrm{T}}
\newcommand{\rD}{\mathrm{D}}
\newcommand{\rG}{\mathrm{G}}
\newcommand{\rPG}{\mathrm{PG}}

\newcommand{\y}{\boldsymbol{y}}

\begin{document}
%
\title{
Turbo-Equalization Using Partial Gaussian Approximation}

\author{ Chuanzong~Zhang, Zhongyong~Wang, Carles~Navarro~Manch\'on, Peng~Sun, Qinghua Guo
        and~Bernard~Henri~Fleury,~\IEEEmembership{Senior member,~IEEE}
\thanks{This work is supported by the National Natural Science
Foundation of China (NSFC 61571402, NSFC U1204607, NSFC 61201251).
}
\thanks{C. Zhang is with the School of Information Engineering, Zhengzhou University, Zhengzhou 450001, China, and the Department of Electronic Systems, Aalborg University, Aalborg 9220, Denmark (e-mail: ieczzhang@gmail.com).}
\thanks{Z. Wang and P. Sun are with the School of Information Engineering, Zhengzhou University, Zhengzhou 450001, China (e-mail: iezywang@zzu.edu.cn; iepengsun@gmail.com).}
\thanks{C. Navarro Manch\'on and B. H. Fleury are with the Department of Electronic Systems, Aalborg University, Aalborg 9220, Denmark (e-mail: cnm@es.aau.dk; fleury@es.aau.dk).}
\thanks{Q. Guo is with the School of Electrical, Computer and Telecommunications Engineering, University of Wollongong, Wollongong, NSW 2522, Australia, and also with the School of Electrical, Electronic and Computer Engineering, the University of Western Australia, Crawley, WA 6009, Australia (e-mail: qguo@uow.edu.au).}
}

\maketitle

\begin{abstract}

This paper deals with turbo-equalization for coded data transmission over intersymbol interference (ISI) channels. We propose a message-passing algorithm that uses the expectation-propagation rule to convert messages passed from the demodulator-decoder to the equalizer and computes messages returned by the equalizer by using a partial Gaussian approximation (PGA). Results from Monte Carlo simulations show that this approach leads to a significant performance improvement compared to state-of-the-art turbo-equalizers and allows for trading performance with complexity. We exploit the specific structure of the ISI channel model to significantly reduce the complexity of the PGA compared to that considered in the initial paper proposing the method.

\end{abstract}

\begin{IEEEkeywords}
Turbo equalization, partial Gaussian approximation, message-passing.
\end{IEEEkeywords}

\IEEEpeerreviewmaketitle

 \section{Introduction}\label{Sec:intro}

\IEEEPARstart{H}{istorically}, turbo equalization of coded data transmission across a known inter-symbol interference (ISI) channel found its inspiration from turbo-decoding of turbo-codes, see \cite{Tuchler2002_1} and references therein. Since its introduction turbo equalization has prevailed over more traditional equalization techniques available at that time due to its tremendous performance gain. Turbo-equalization is a collective name for joint data decoding and channel equalization algorithms that pass messages iteratively along the edges of a factor graph representing the probabilistic model of the considered transmission system. The most prominent message-passing algorithm --  inherited from turbo-decoding of turbo-codes -- is the sum-product algorithm~\cite{Loeliger2004}, which is also known as belief propagation (BP)
\cite{Pearl982reverendbayes}.

Two different factor graphs~\cite{Loeliger2004}
representing the ISI channel can be drawn, which lead to different message-passing algorithms for equalization, see \cite{Kurkoski2002} and \cite{Colavolpe2005} for more details. In this letter, we use the one that exhibits a tree structure \cite{Kurkoski2002}. This factor graph explicitly represents the channel state evolution,
see Fig.~\ref{fig:FG}. Applying BP on this graph yields BCJR-like equalization algorithms \cite{Kurkoski2002}. The complexity of these algorithms scales exponentially with the modulation order and the channel memory.
Proposed solutions that circumvent this complexity problem convert the discrete messages returned by the demodulator-decoder into Gaussian functions that are passed as messages to the equalizer \cite{Hu2006,Guo2008,Sun2015}. The complexity reduction results from the fact that the equalizer then processes Gaussian messages.
The conversion can be done in two ways: either directly by matching the first and second moments of any of these discrete messages \cite{Guo2008}, or indirectly by using the formal rule of expectation propagation~(EP)~\cite{Minka2001}: first
a Gaussian approximation matching the first and second moments of the belief of the channel symbol node is computed from which the Gaussian message is then obtained  \cite{Hu2006}, \cite{Sun2015}.
Numerical studies have shown that the latter conversion leads to better BER performance \cite{Hu2006}, \cite{Sun2015}.

Inspired by the partial Gaussian approximation (PGA) proposed in \cite{Guo2014} we modify the messages returned from the equalizer and passed to the demodulator-decoder in \cite{Sun2015}. The messages returned by the equalizer in \cite{Sun2015} are computed from the above Gaussian-converted messages from the demodulator-decoder. By contrast, to equalize one channel symbol the new equalizer combines discrete messages from the demodulator-decoder for the symbols strongly interfering with said symbol and Gaussian-converted messages for the weakly interfering symbols. The reported simulation results show that doing so leads to a significant performance improvement compared to the turbo-equalizers in \cite{Guo2008}, \cite{Guo2014} and \cite{Sun2015}. Finally, our turbo-equalizer allows for trading complexity with performance by varying the set of symbols that are considered as strong interferers.

Our turbo-equalizer differs from the PGA-based one in \cite{Guo2014} in two respects. First, in the former the conversion of the discrete messages returned by the demodulator-decoder into Gaussian functions is done using the formal EP rule, while it is performed by direct conversion of the discrete messages in the latter. Secondly, due to the particular structure of the ISI channel model, the messages returned by our equalizer can be computed from the Gaussian messages passed to it
in a simple way.  This leads to a significant complexity reduction compared to the turbo-equalizer in \cite{Guo2014}.

\textit{Notation}- For a natural number $N$,  we write $\subs{N}=\{1,\ldots, N\}$. Boldface lowercase and uppercase letters denote vectors and matrices, respectively. The identity matrix of size $M$ is represented by $\boldsymbol{I}_M$. Superscript $(\cdot){\tra}$ designates transposition of a vector or matrix. We write $\gpdf(\x;\m,\V)$ for the pdf of a multivariate Gaussian distribution with mean vector $\m$ and covariance matrix $\V$. Depending on the context $\delta(\cdot)$ denotes either the Dirac delta function or the Kronecker delta. The relation $f(\cdot)=cg(\cdot)$ for some positive constant $c$ is written as $f(\cdot)\propto g(\cdot)$. The notations $\sum_{\x\setminus \y}f(\x)$ and $\int f(\x) \dd(\x\setminus \y)$ denote respectively the partial summation and partial integration of the function $f(\x)$ with respect to all entries of the vector $\x$ except those entries common to $\x$ and $\y$.

\section{System Model}\label{Sec:SysMod}
The vector $\bb=[b_1,\dots,b_{K}]^ {\textrm{T}}$ of information bits is encoded and interleaved, yielding the codeword $\cc=[\cc_1\tra,\dots,\cc_{N}\tra ]^ {\textrm{T}}$ with $\cc_i=[c^1_i,\cdots,c_i^{Q}]\tra$.  The coded bits are then mapped onto a $Q$-order
modulation alphabet $\MA\subseteq\mathbb R$ \footnote{For simplicity we consider a real baseband model. The extension to a complex model is straightforward.},
resulting in the vector of symbols $\x=[x_1,\dots,x_{N}]^ {\textrm{T}}\in\MA^N$. These symbols are transmitted over a linear, time-invariant, frequency-selective channel corrupted by additive white Gaussian noise (AWGN). The received vector $\rr=[r_1,\dots,r_{N+L-1} ]^ {\textrm{T}}$ has entries
\begin{eqnarray}
r_i= \sum_{l=0}^{L-1}h_lx_{i-l}+n_i=\h^ {\textrm{T}}\s_i+n_i\label{eq:signalmodel}, \quad i\in\subs{N+L-1}
\end{eqnarray}
where $\s_i=[x_{i-L+1},\dots,x_i]^ {\textrm{T}}$ with $x_{i}=0$  for  $i<1$ and $i>N$, $\h=[h_{L-1},\dots,h_0 ]^ {\textrm{T}}$ represents the channel impulse response, and $\n=[n_1,\dots,n_{N+L-1} ]^ {\textrm{T}}$ is a white noise vector with component variance $\sigma^2$.

\subsection{Probabilistic Model and Factor Graph}\label{Sec:model}

The posterior probability mass function (pmf)  of vectors $\bb$, $\cc$, $\x$ and $\s$ given the received signal $\rr$ reads
\begin{align}
{p\left( \bb,\cc,\x,\s|\rr\right)}
& \propto  \prod_{k=1}^{K}f_{\rB_k}\left(b_k \right) \times f_\rC\left(\cc,\bb \right)  \nonumber \\
& \times  \prod_{i=1}^{N}f_{\rO_i}\left(r_i,\s_i \right)f_{\rT_i}\left(\s_i,\s_{i-1},x_i \right) f_{\md_i}\left(x_i,\cc_i \right) \nonumber\\
& \times  \prod_{i=N+1}^{N+L-1}f_{\rO_i}\left(r_i,\s_i \right)f_{\rT_i}\left(\s_i,\s_{i-1},0\right)
\label{eq:factorization}
\end{align}
where $f_{\rB_k}(b_k)$  is the uniform prior pmf of the $k$th information bit, $f_\rC(\cc,\bb) $ stands for the coding and interleaving constraints, $f_{\rO_i}(r_i,\s_i )\triangleq p(r_i|\s_i )= \gpdf(r_i;\h^ {\textrm{T}}\s_i,\sigma^2 )$ denotes the likelihood of $\s_i$, 
and $f_{\md_i}(x_i,\cc_i)$ represents the modulation mapping.
Finally, $f_{\rT_i}\left(\s_i,\s_{i-1},x_i \right)$ expresses the deterministic relationship between $\s_i$, $\s_{i-1}$ and $x_i$, i.e.,
\begin{eqnarray}
f_{\rT_i}\left(\s_i,\s_{i-1},x_i \right)=\delta(\G\s_{i-1}+\e x_i-\s_i)\label{eq:G and f}
\end{eqnarray}
with the $L\times L$ matrix $\G=\left[
\begin{array}{l}
 \0\,\,\,\,\,\, \I_{L-1};\
 0                \,\,\,\, \,\,\0^ {\textrm{T}}
\end{array}
\right]$,
$\e=\left[
\begin{array}{l}
 \0 ;
\ 1
\end{array}
\right] $ and $\0$ being a zero column vector of length $L-1$.

\begin{figure}[!t]
\centering
\includegraphics[width=0.4\textwidth]{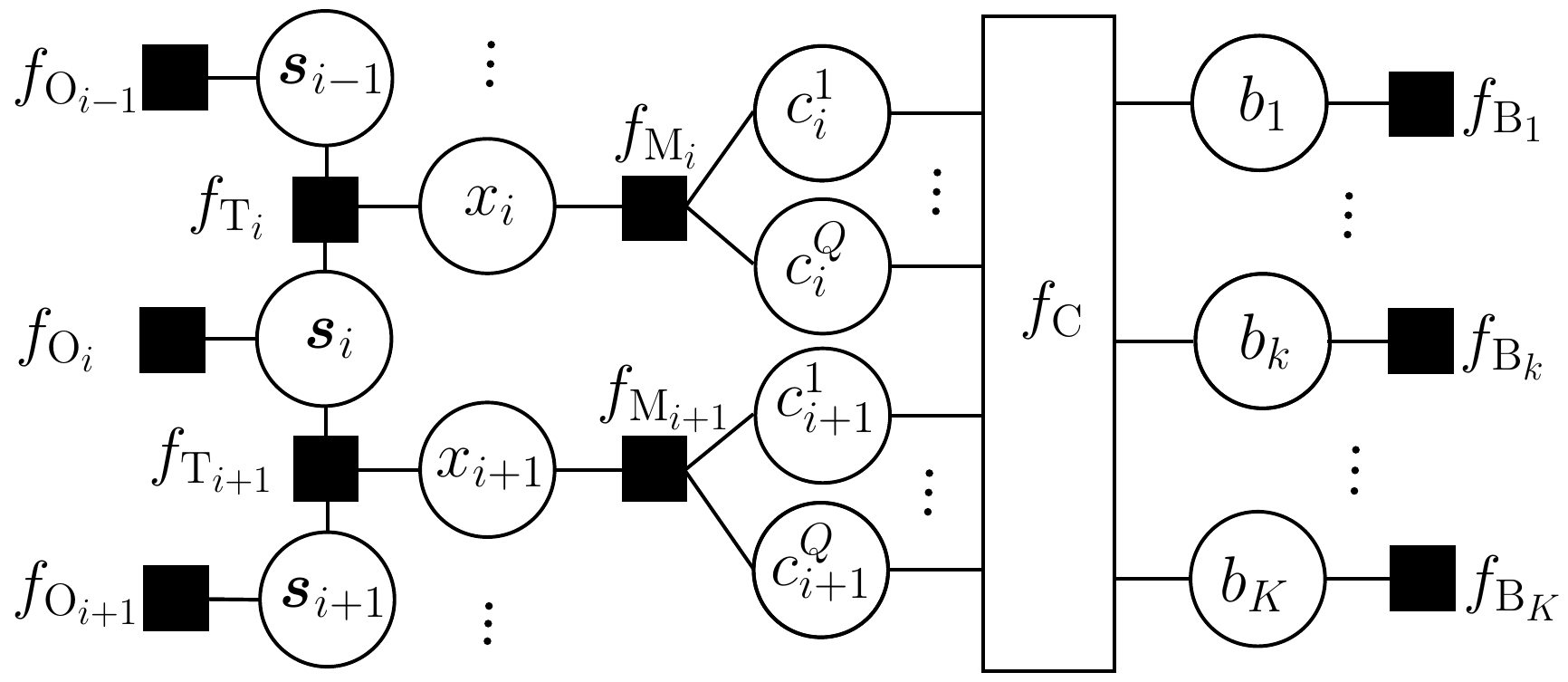}
\caption{Factor graph representing the probabilistic model~\eqref{eq:factorization}.}
\label{fig:FG}
\end{figure}

Fig.~\ref{fig:FG} depicts the factor graph \cite{Kschischang2001} representing the factorization of the posterior pmf in (\ref{eq:factorization}). The factorization and its graph will be used as the baseline for the derivation of the turbo-equalizer described in Section~\ref{Sec:Design}. To ease the subsequent discussions we identify two subgraphs. The channel subgraph includes the nodes of the channel symbols $x_i$, $i\in\subs{N}$ and all factor nodes, variable nodes and edges ``to the left'' of these symbol nodes. The transmitter subgraph includes the channel symbol nodes and all factor nodes, variable nodes, and edges ``to the right'' of these symbol nodes.

\section{Design of the Iterative Receiver}\label{Sec:Design}
In a nutshell, we obtain the new turbo-equalizer by replacing the messages passed from the equalizer to the demodulator-decoder, i.e. from nodes $f_{\rT_i}$ to nodes $x_i$, $i\in\subs{N}$, in the turbo-equalizer in \cite{Sun2015} by messages computed using the PGA approach in \cite{Guo2014}.
The next two subsections describe the messages computed in the new turbo-equalizer. The last subsection sketches the scheduling of these messages.

\subsection{Equalization and Demodulation-decoding}\label{sec:MessagesInSubgraphs}

Equalization and demodulation-decoding are implemented by passing messages along the edges of the  channel subgraph and the transmitter subgraph respectively. Unless otherwise stated, these messages are computed using the BP rule \cite{Pearl982reverendbayes}.

\subsubsection{Demodulation-decoding}\label{sec:PartTS}
The variables in the transmitter subgraph are discrete and so are the computed messages and beliefs. Decoding of the convolution code is done using the BCJR algorithm, an instance of BP. The messages from the modulator nodes to the channel symbol nodes are of the form
$m_{f_{\md_i}\to x_i}\!(x_i)\propto
\sum_{x\in\MA}\beta_{x} \delta\left(x_i-x\right)$ with $\beta_x\geq 0$, $x_i\in\MA$, $i\in\subs{N}$.

\subsubsection{Equalization}\label{Sec:Up-Down}
The latent variables $\s_{i}$, $i\in\subs{N}$ in the channel subgraph are approximated as Gaussian variables. Since the channel is linear and noise is additive and Gaussian, the messages and beliefs are Gaussian functions. We write for the belief of node $\s_{i}$ $(i\in\subs{N})$,
\begin{eqnarray}\label{eq:belief}
b^{\GA}(\s_{i}) 
\propto \gpdf(\s_i;\m_{\s_i},\V_{\!\!\s_i}).
\label{eq:fGB}
\end{eqnarray}
The computation of this belief is given in \cite{Guo2008} and \cite[Eq. (28)]{Sun2015}.

\subsection{Messages Exchanged Between the Equalizer and the Demodulator-decoder}\label{S:Conv}
\subsubsection*{\hspace*{-3ex} Demodulator-decoder (D)$\,\,\rightarrow\,\,$Equalizer (E)}
The EP rule~\cite{Minka2001}
is used to convert the discrete messages $m_{f_{\md_i}\to x_i}\!(x_i)$,
$i\in\subs{N}$ into Gaussian messages~\cite{Hu2006},~\cite[Eq. (29)]{Sun2015}:
\begin{eqnarray}
m^\GA_{f_{\md_i}\to x_i}\!(x_i) &=&\frac{\textrm{Proj}_{\mathcal{G}}
[ m_{f_{\md_i}\to x_i}\!(x_i) n^\GA_{x_i\to f_{\md_i}} \!(x_i)
] }{n^\GA_{x_i\to f_{\md_i}}\!(x_i)} \nonumber\\
&\propto& \gpdf( x_i;
m_{x_i},v_{x_i}),
\quad i\in\subs{N}.
\label{eq:fMitoxi}
\end{eqnarray}
For a pdf $b(z)$, $\textrm{Proj}_{\cal{G}}[b(z)]= \argmin_{b'(z) \in \mathcal{G}} \mathrm{D} \left(b(z) || b'(z)\right)$, with $\mathrm{D}(\cdot||\cdot)$ denoting the Kullback-Leibler divergence and $\mathcal{G}$ being the family of Gaussian pdfs.
The parameters
$m_{x_i}$ 
and $v_{x_i}$
 in (\ref{eq:fMitoxi}) are given by~\cite[Eq. (10) \& (11)]{Sun2015}.
With this conversion, Gaussian messages $n^{\GA}_{x_i\to f_{\rT_i}}\!(x_i)= m^{\GA}_{f_{\md_i}\to x_i}\!(x_i)$, $i\in\subs{N}$ are passed to the equalizer.

\subsubsection*{\hspace*{-2.5ex} E$\,\,\rightarrow\,\,$D}\label{Sec:output}
This is where the new turbo-equalizer differs from the one described in \cite{Hu2006}, \cite{Sun2015}.

In \cite{Hu2006}, \cite{Sun2015} the Gaussian messages from $f_{\rT_i}$ to $x_i$, $i\in\subs{N}$ are converted into discrete messages
\footnote{Strictly speaking, the message ${m}_{f_{\rT_i}\to x_i}(x_i)$ in (\ref{eq:fDisc}) is the restriction of ${m}_{f_{\rT_i}\to x_i}^{\GA}(x_i)$ to $\MA$.}
\begin{eqnarray}
{m}_{f_{\rT_i}\to x_i}(x_i) \propto {m}_{f_{\rT_i}\to x_i}^{\GA}(x_i)
,\quad  x_i \in \MA, i\in\subs{N}.
\label{eq:fDisc}
\end{eqnarray}
The discrete messages ${n}_{x_i\to f_{\md_i}}(x_i)={m}_{f_{\rT_i}\to x_i}(x_i)$, $i\in\subs{N}$ are then passed to the demodulator-decoder.

Consider a specific symbol $x_i$ ($i\in\subs{N}$). Clearly the computation of ${m}_{f_{\rT_i}\to x_i}(x_i)$ using (\ref{eq:fDisc}) makes use of the Gaussian approximation of the messages from  the other symbols, i.e.~${n}_{x_j\to f_{\rT_j}}^{\GA}(x_j)$, $j\in\subs{N}\setminus\{i\}$ by the conversion (\ref{eq:fMitoxi}). The idea is to use the original discrete messages rather than their Gaussian approximation for a selected subset of channel symbols which significantly interfere with $x_i$. It is inspired from the PGA proposed in~\cite{Guo2014}.

\begin{figure*}[!t]
\begin{align}
\lefteqn{
b^\rG(\s_{i'}) =
m^\rG_{f_{\rT_{i'+1}}\!\to \s_{i'}}\!(\s_{i'})\,
m^\rG_{f_{\rO_{i'}}\to \s_{i'}}\!(\s_{i'})
\left[
\prod_{l=0}^{L-1}
n^\rG_{x_{i'-l} \to f_{\rT_{i'-l}}}\!(x_{i'-l})
\right]
 } \hspace*{10ex} \nonumber \\
 &
\times
\int
\left[
\prod_{l=1}^{L-1}
m^\rG_{f_{\rO_{i'-l}}\!\to \s_{i'-l}}\!(\s_{i'-l})
\right]
n^\rG_{\s_{{i'}-L}\to f_{\rT_{{i'}-L+1} }}\!(\s_{{i'}-L})
\dd\s_{i'-L}
\label{eq:EBE}
\end{align}
\end{figure*}

First we identify those channel symbols ``significantly'' interfering with symbol $x_i$.
Let $q_k=\sum_{l=0}^{L-1}h_lh_{l+k}$, $k\in\mathbb Z$ with $h_l=0$ whenever $l\in\mathbb Z\setminus\{0,\ldots,L-1\}$ be the autocorrelation function of the channel impulse response.
Define $\mathbb K_\rho=\{k\in  \{-(L-1),\ldots,L-1\}: \vert q_k \vert > \rho q_0  \}$ the set of lag indices at which the magnitude of the autocorrelation function is larger than $\rho q_0$, $\rho\in[0,1)$. Then $\mathbb I_i^\rD=\{i+k:k\in\mathbb K_\rho\}\subseteq  \mathbb I_i=\{i-(L-1),\dots,i+L-1\}$ contains the indices of the modulation symbol $x_i$ and those symbols that interfere with $x_i$ at correlation level $\rho$. We collect these symbols in the $M$-dimensional vector $\x_i^\rD=[x_j:j\in \mathbb I_{i}^\rD]^{\textrm{T}}$, with $M=|\mathbb K_\rho|$. 
We assume that $\bar{k}=\max \mathbb K_\rho$ fulfills $1+2\bar{k}\leq L$. Then we can readily show that all entries in $\x_i^\rD$
are components of $\s_{i'}$ whenever $i+\bar{k}\leq i' \leq i+(L-1)-\bar{k}$. Notice that the assumption on $\bar{k}$ guarantees that $i+\bar{k} \leq i+(L-1)-\bar{k}$.

With the above definitions we can now specify the message from $f_{\rT_i}$ to $x_i$:
\begin{align}
m^\rPG_{f_{\rT_i}\to x_i}\!(x_i)= &
\sum_{\x_{i}^\rD\setminus x_i}
\frac{
\prod_{\kappa\in \mathbb{I}_i^\rD \setminus i}n_{x_\kappa \to f_{\rT_\kappa}}\!(x_\kappa)
}{
\prod_{k\in \mathbb{I}_i^\rD}n^\rG_{x_k\to f_{\rT_k}}\!(x_k)
}
b_{i'}^\GA(\x_i^\rD)
\label{eq:BPfGtoX}
\end{align}
where $n_{x_\kappa \to f_{\rT_\kappa}}\!(x_\kappa)=m_{f_{\md_\kappa}\to x_\kappa}\!(x_\kappa)$ and $b_{i'}^\GA(\x_i^\rD)=\int  {b}^\GA(\s_{i'})
\dd(\s_{i'}\!\!\setminus\!\x_i^\rD)$ with ${b}^\GA(\s_{i'})$ given in (\ref{eq:fGB}). The latter term is the belief of $\x_i^\rD$ obtained by marginalization of the belief $b^\GA(\s_{i'})$. The index $i'$ in $b_{i'}^\GA(\x_i^\rD)$ indicates that this belief depends on the time instant $i'$, $i+\bar{k}\leq i' \leq i+(L-1)-\bar{k}$.
Notice that the selection $i'=i+\bar{k}$ minimizes the time instant ahead of $i$ to wait for computing ${m}_{f_{\rT_i}\to x_i}(x_i)$. The derivation of (\ref{eq:BPfGtoX}) is provided in the appendix.

All Gaussian functions occurring in (\ref{eq:BPfGtoX}) combine as
\begin{align}
\left[
\prod_{\kappa\in \mathbb{I}_i^\rD}n^\rG_{x_\kappa \to f_{\rT_\kappa}}\!(x_\kappa)
\right]^{-1}
\!\!\!
b_{i'}^\GA(\x_i^\rD)
 & \propto \gpdf (\x_i^\rD;\m^e_{\x_i^\rD},\V^e_{\x_i^\rD})
\label{eq:gp}
\end{align}
with
\begin{align}\small
\V_{\x_i^\rD}^e &= \left[(\boldsymbol{P}_{i'} \V_{\s_{i'}}\boldsymbol{P}_{i'}\tra)^{-1}-(\V_{\x_i^\rD})^{-1}\right]^{-1}\nonumber\\
\m_{\x_i^\rD}^e&=\V_{\x_i^\rD}^e\left[(\boldsymbol{P}_{i'} \V_{\s_{i'}}\boldsymbol{P}_{i'}\tra)^{-1}\boldsymbol{P}_{i'}\m_{\s_{i'}}- (\V_{\x_i^\rD})^{-1}\m_{\x_i^\rD}\right]. \nonumber
\end{align}
In these expressions, the $M\times L$ selection matrix $\boldsymbol P_{i'}$ extracts the vector $\x_{i}^\rD$ from $\s_{i'}$, i.e.\
$\x_{i}^\rD=\boldsymbol{P}_{i'}\s_{i'}$.
The entries of the vector $\m_{\x_i^\rD}$ and the diagonal entries of the diagonal matrix $\V_{\x_i^\rD}$
are the first moments $m_{x_\kappa}$ and the second central moments $v_{x_\kappa}$ respectively
of the messages $n^\rG_{x_\kappa \to f_{\rT_\kappa}}\!(x_\kappa)$, $\kappa\in \mathbb{I}_i^\rD$.
Inserting (\ref{eq:gp}) into (\ref{eq:BPfGtoX}) yields
the PGA-based messages
\begin{align}
{m}^\rPG_{f_{\rT_i}\to x_i}(x_i)= & \!
\sum_{\x_{i}^\rD\setminus x_i}
\!\!
\gpdf (\x_i^\rD;\m^e_{\x_i^\rD},\V^e_{\x_i^\rD})
\!\!\!
\prod_{\kappa\in \mathbb{I}_i^\rD \setminus i}
\!\!n_{x_\kappa \to f_{\rT_\kappa}}\!(x_\kappa),  \nonumber \\
 & \quad\quad\quad\quad\quad\quad\quad\quad\quad\quad\quad\quad i\in\subs{N}
\label{eq:BPfGtoXF}
\end{align}
that
replace the messages ${m}^\GA_{f_{\rT_i}\to x_i}(x_i)$, $i\in\subs{N}$ in (\ref{eq:fDisc}).


\subsection{Messages Scheduling}
The turbo-equalizer implements the following scheduling:
\begin{enumerate}[S1:]
\item {\em Initialization:}
 $n_{x_i\to f_{\rT_i}}\!(x_i)\propto 1$ and $ n^\GA_{x_i\to f_{\rT_i}}\!(x_i)=\mathcal N(x_i;0,1)$, $i\in\subs{N}$. 
\item {\em Equalization:} The messages  $m^\rG_{f_{\rT_i}\to \s_{i} }\!(\s_{i})$ and $n^\rG_{\s_{i}\to f_{\rT_{i+1}}}\!(\s_{i})$,  $i\in\subs{N+L-1}$ are recursively computed using (12) and  (15) respectively in~\cite{Sun2015}. In parallel, the messages $m^\rG_{f_{\rT_i}\to \s_{i-1} }\!(\s_{i-1})$ and $n^\rG_{\s_{i-1}\to f_{\rT_{i-1}}}\!(\s_{i})$, $i\in\{N+L-1,\dots,1\}$ are recursively calculated  from (18) and  (21) respectively in~\cite{Sun2015}. Finally, the beliefs $b^{\GA}(\s_{i})$, $i\in [N]$ (see (\ref{eq:fGB})) are obtained by (28) in~\cite{Sun2015}.

\item {\em E$\,\,\rightarrow\,\,$D:} The messages $m^\rPG_{f_{\rT_i}\to x_i}\!(x_i)$, $i\in\subs{N}$ are obtained from \eqref{eq:gp} and (\ref{eq:BPfGtoXF}).
\item {\em Demodulation-decoding:} The  messages $m^\rPG_{f_{\rT_i}\to x_i}\!(x_i)$, $i\in\subs{N}$ are passed to the demodulator. The BCJR algorithm, an instance of BP, is run in the decoder, yielding the discrete messages $n_{x_i\to f_{\rT_i}}\!(x_i)=m_{f_{\md_i}\to x_i }\!(x_i)$, $i\in\subs{N}$.
\item {\em D$\,\,\rightarrow\,\,$E:} The Gaussian messages $n^\GA_{x_i\to f_{\rT_i} }\!(x_i)=m^\GA_{f_{\md_i}\to x_i }\!(x_i)$, $i\in\subs{N}$ are updated using (\ref{eq:fMitoxi}).
\end{enumerate}
Steps S2--S5 constitute an iteration that is repeated until a maximum number of iterations is reached.

\section{Analysis, Performance and Complexity}\label{Sec:sim}

\subsection{Comparison with Existing Turbo-equalizers}
We compare the performance of the new turbo-equalizer (we denote it as BP-EP-PGA)
with that of three other turbo-equalizers published in the literature by means of Monte Carlo simulations:
(1) BP-EP: the combined BP-EP algorithm  in \cite{Sun2015};
(2) BP-PGA: an implementation of the PGA algorithm in \cite{Guo2014} for the equalization of ISI channels;
(3) BP-GA: the LMMSE-based turbo-equalizer, which is equivalent to Gaussian-approximated BP \cite{Guo2008}.
In our implementation BP-PGA is obtained from BP-EP-PGA by substituting the EP rule (\ref{eq:fMitoxi}) with a direct Gaussian approximation of the discrete messages from $f_{\md_i}$ to $x_i$, $i\in\subs{N}$.
All four turbo-equalizers solely differ in the types of messages exchanged between the equalizer and the demodulator-decoder. The table below
reports these distinctive features.
%
\begin{center}\small
\begin{tabular}{| l | c | c |} \hline
Turbo-equalizer & {\em D$\,\,\rightarrow\,\,$E} & {\em E$\,\,\rightarrow\,\,$D} \\ \hline
BP-GA \cite{Guo2008} & Direct conversion & GA \\ \hline
BP-EP \cite{Sun2015} & EP-rule & GA \\ \hline
BP-PGA \cite{Guo2014} & Direct conversion & PGA \\ \hline
BP-EP-PGA (new)& EP-rule & PGA \\ \hline
\end{tabular}
\end{center}
As the selected threshold $\rho$ in BP-EP-PGA approaches $1$, $\mathbb{I}_i^\rD$ typically shrinks to the singleton
$\{i\}$ ($M=1$). With this configuration, the messages ${m}^\rPG_{f_{\rT_i}\to x_i}$, $i\in\subs{N}$ in (\ref{eq:BPfGtoXF}) coincide with the messages ${m}^\GA_{f_{\rT_i}\to x_i}$, $i\in\subs{N}$ in (\ref{eq:fDisc}) and, consequently, BP-EP-PGA and BP-EP become equivalent. 
Notice that both schemes compute the same messages in stage S2-Equalization. They solely differ in S5-{\em D}$\,\,\rightarrow\,\,${\em E}.

\subsection{Computational Complexity}\label{sec:CompComp}
The complexity of the PGA algorithm in~\cite{Guo2014}, which was designed for generic channel matrices, is $\mathcal{O}(N^2+M^2Q^{M})$ per symbol. The main contribution to the complexity of the BP-EP-PGA is at (\ref{eq:belief}), which requires $L\times L$ matrix inversions (see \cite[Eq. (28)]{Sun2015}), and at (\ref{eq:BPfGtoX}). Thus, the complexity is $\mathcal{O}(L^3+M^2Q^{M})$ per symbol. The complexity reduction method described in \cite[Subsec.~IV.C]{Sun2015} can, however, also be applied to BP-PGA and BP-EP-PGA. Since the beliefs $b_{i'}^\GA(\x_i^\rD)$ (see (\ref{eq:BPfGtoX})) are obtained from the beliefs $b^\GA(\s_{i'})$, $i\in\subs{N}$ (\ref{eq:belief}) needs only be computed once every $(L-M+1)$ symbols
when $\mathbb K_\rho=\{-\bar{k},\ldots,\bar{k}\}$ ($M=1+2\bar{k}$).
In this case, the complexity of BP-PGA and BP-EP-PGA is $\mathcal{O}(L^3/(L-M+1)+M^2Q^{M})$ per symbol.

\subsection{Numerical Assessment}
We compare the BER performance of the four above turbo-equalizers and a receiver designed for and operating in a non-dispersive AWGN channel.

A sequence of $2048$ information bits is encoded using a $1/2$ rate convolutional code with generator polynomials $(23,35)_8$. The coded bits are interleaved and then mapped onto BPSK symbols ($\MA=\{-1,+1\}$), which are transmitted over a severely distorted ISI channel with impulse response $\h=[0.227\  0.460\  0.668\  0.460\  0.227]\tra$.
The BER performance is evaluated after $30$ turbo-equalization iterations. 
For  BP-PGA and BP-EP-PGA, we set $\rho$ so that $M=3$.

\begin{figure}[!t]
\centering
\includegraphics[width=0.37\textwidth]{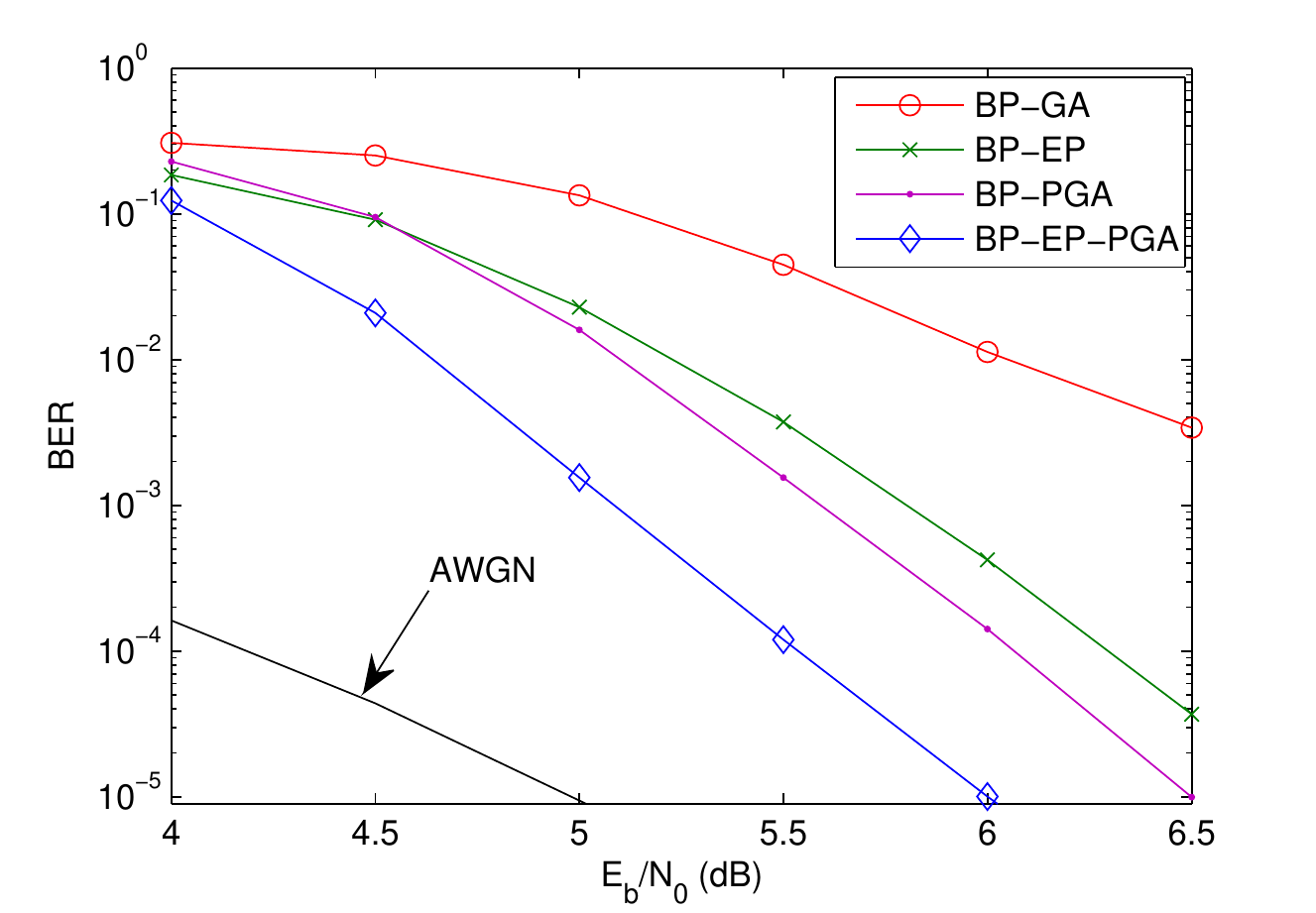}
\caption{BER performance of the considered turbo-equalizers.
}\label{fig:snr_ber}
\end{figure}

The results are depicted in Fig.~\ref{fig:snr_ber}. We observe  a remarkable performance improvement of BP-EP-PGA compared to the other turbo-equalizers. We attribute this improvement to the fact that BP-EP-PGA combines the advantages from both BP-EP and BP-PGA. Firstly, implementing the EP-based conversion (\ref{eq:fMitoxi}) instead of a direct conversion of the discrete messages from $f_{\md_i}$ to $x_i$, $i\in\subs{N}$
provides an advantage over BP-GA and BP-PGA.
Secondly, implementing (\ref{eq:BPfGtoXF}) leads to better performance than when computing the right-hand messages in (\ref{eq:fDisc}) at the expense of a complexity increase, as can be seen by comparing BP-EP and BP-EP-PGA. Since BP-EP can be seen as an instance of our proposed BP-EP-PGA with the setting $M=1$, we conclude that the tuning of the parameter $M$ (or equivalently $\rho$) allows for trading performance and computational complexity in the receiver.

\appendix
\label{sec:PartEP2}

We compute (\ref{eq:BPfGtoX}) for a specific symbol $x_i$ ($i\in\subs{N}$). Select $\s_{i'}$ with $i'$ satisfying $i+\bar{k}\leq i' \leq i+(L-1)-\bar{k}$, see {\em E$\,\,\rightarrow\,\,$D} in Subsection~\ref{S:Conv}. By applying the BP rule we obtain for the Gaussian belief of $\s_{i'}$
\begin{eqnarray}
b^\rG(\s_{i'}) =
m^\rG_{f_{\rT_{i'}}\!\to \s_{i'}}\!(\s_{i'})
m^\rG_{f_{\rO_{i'}}\to \s_{i'}}\!(\s_{i'})m^\rG_{f_{\rT_{i'+1}}\!\to \s_{i'}}\!(\s_{i'}).
\label{eq:BS}
\end{eqnarray}
To compute $m^\rG_{f_{\rT_{i'}}\!\to \s_{i'}}\!(\s_{i'})$ we use the BP rule in a forward recursion along the variable and factor nodes $f_{\rT_{i'-L+1}},
\s_{i'-L+1},\ldots,\s_{i'-1},f_{\rT_{i'}}$. Doing so and inserting in (\ref{eq:BS}) yields the expression in (\ref{eq:EBE}).
Notice that the product in the first pair of brackets and the integral are functions of $\s_i=[x_{i-L+1},\dots,x_i]^ {\textrm{T}}$.
From the choice of $i'$, the entries of $\x_i^\rD=[x_j:j\in \mathbb I_{i}^\rD]^{\textrm{T}}$ are also entries of $\s_{i'}$, see {\em E$\,\,\rightarrow\,\,$D} in Subsection~\ref{S:Conv}. Thus, the product in the first bracket in (\ref{eq:EBE}) contains as factors the messages $n^\rG_{x_{j} \to f_{\rT_{j}}}\!(x_{j})$, $j\in\mathbb I_i^\rD$. We implement a PGA by substituting these messages with their discrete counterparts $n_{x_{j} \to f_{\rT_{j}}}\!(x_{j})$, $j\in\mathbb I_i^\rD$. This substitution can be formally expressed as
\begin{align}
b^\rPG(\s_{i'}) &=
\prod_{\kappa\in \mathbb{I}_i^\rD}
\frac{
n_{x_\kappa \to f_{\rT_\kappa}}\!(x_\kappa)
}{
n^\rG_{x_\kappa \to f_{\rT_\kappa}}\!(x_\kappa)
}
\,b^\GA(\s_{i'})
\label{eq:beliefsi}.
\end{align}
By using the marginalization constraint of BP we can write
\begin{eqnarray}	
m^\rPG_{f_{\rT_i}\to x_i}\!(x_i)n_{x_i\to f_{\rT_i}\!}(x_i) 
= \sum_{\x_{i}^\rD\setminus x_i}
\!\int b^\rPG(\s_{i'})\dd(\s_{i'}\!\!\setminus\! \x^\rD_i).
\label{eq:BPfGtoX2}
\end{eqnarray}
Notice that the right-hand term is a marginal belief of $x_i$. Solving for $m^\rPG_{f_{\rT_i}\to x_i}\!(x_i)$ in (\ref{eq:BPfGtoX2}) yields (\ref{eq:BPfGtoX}).

%

\begin{thebibliography}{99}
\bibitem{Tuchler2002_1} M. T\"uchler, R. Koetter, and A. Singer, ``Turbo equalization: Principles
and new results," IEEE Trans. Commun., vol. 50, pp. 754 -- 767, May
2002.
\bibitem{Loeliger2004} H.-A. Loeliger, ``An introduction to factor graphs," IEEE Trans. Signal
Processing, pp. 28 -- 40, Jan. 2004.
\bibitem{Pearl982reverendbayes} J. Pearl, ``Reverend Bayes on inference engines: a distributed hierarchical
approach," in in Proceedings of the National Conference on Artificial
Intelligence, 1982, pp. 133 -- 136.
\bibitem{Kurkoski2002} B. Kurkoski, P. Siegel, and J. Wolf, ``Joint message-passing decoding
of ldpc codes and partial-response channels," IEEE Transactions on Information Theory, vol. 48, no. 6, pp. 1410 -- 1422, Jun 2002.
\bibitem{Colavolpe2005} G. Colavolpe and G. Germi, ``On the application of factor graphs and the
sum-product algorithm to ISI channels," IEEE Trans. Commun., vol. 53,
no. 5, pp. 818 -- 825, Apr. 2005.
\bibitem{Hu2006} J. Hu, H.-A. Loeliger, J. Dauwels, and F. Kschischang, ``A general
computation rule for lossy summaries/messages with examples from
equalization," in Proc. 44th Allerton Conf. Communication, Control, and
Computing, Sep. 2006, pp. 27 -- 29.
\bibitem{Guo2008} Q. Guo and L. Ping, ``LMMSE turbo equalization based on factor
graphs," IEEE J. Select. Areas Commun., vol. 26, no. 2, pp. 311 -- 319,
Feb. 2008.
\bibitem{Sun2015} P. Sun, C. Zhang, Z. Wang, C. N. Manch\'on, and B. H. Fleury, ``Iterative
receiver design for ISI channels using combined belief- and expectation- propagation,"
IEEE Signal Processing Lett., vol. 22, no. 10, pp. 1733 --
1737, Oct. 2015.
\bibitem{Minka2001} T. P. Minka, ``Expectation propagation for approximate Bayesian inference,"
in Proceedings of the Seventeenth Conf. on Uncertainty in
Artificial Intelligence, 2001, pp. 362 -- 369.
\bibitem{Guo2014} Q. Guo, D. Huang, L. Ping, S. Nordholm, J. Xi, and P. Li, ``Soft-in
soft-out detection using partial Gaussian approximation," IEEE Access,
vol. 2, pp. 427 -- 436, May 2014.
\bibitem{Kschischang2001} F. Kschischang, B. Frey, and H.-A. Loeliger, ``Factor graphs and the
sum-product algorithm," IEEE Trans. Inform. Theory, vol. 47, no. 2, pp.
498 -- 519, Feb. 2001.
\end{thebibliography}

\end{document}